\documentclass[prl, twocolumn,amsfonts]{revtex4}
\usepackage{amsfonts}
\usepackage{amsmath}
\usepackage{bm}
\usepackage{epsf}

\newcommand{\beq}{\begin{equation}}
\newcommand{\eeq}{\end{equation}}
\newcommand{\beqa}{\begin{eqnarray}}
\newcommand{\eeqa}{\end{eqnarray}}

\newcommand{\ket}[1]{| #1    \rangle }
\newcommand{\bra}[1]{ \langle   #1  | }

\newcommand{\qave}[2]{  \langle #1  | #2  | #1  \rangle }
\newcommand{\mel}[3]{  \langle #1  | #2   | #3  \rangle }
\newcommand{\amp }[2]{ \langle #1 |  #2  \rangle }

\newcommand{\vect}[1]{{\bm{ #1}}}

\newcommand{\xmin}{x_{\!{}_-}}
\newcommand{\pmin}{p_{\!{}_-}}
\newcommand{\xplus}{x_{\!{}_+}}
\newcommand{\pplus}{p_{\!{}_+}}

\newcommand{\hatxmin}{\hat{x}_{\!{}_-}}
\newcommand{\hatpmin}{\hat{p}_{\!{}_-}}
\newcommand{\hatxplus}{\hat{x}_{\!{}_+}}
\newcommand{\hatpplus}{\hat{p}_{\!{}_+}}
\newcommand{\hatxpm}{\hat{x}_{\!{}_\pm}}
\newcommand{\hatppm}{\hat{p}_{\!{}_\pm}}

\begin{document}

%%%%%%%%%%%%%%%%%%%%%%%%%%%%%%%%%%%%%%%%%%%%%%%%%%%%%%%%%%%%%%%%%%%%%%%%
%%%%%%%%%%%%%%%%%%%% Local Definitions %%%%%%%%%%%%%%%%%%%%%%%%%%%%%%%%%

\newcommand{\diracsl}[1]{\not\hspace{-3.0pt}#1}

\newcommand{\pin}{\psi_{\!{}_1}}
\newcommand{\psf}{\psi_{\!{}_2}}
\newcommand{\psv}[1]{\psi_{\!{}_#1}}
\newcommand{\lab}[1]{{}^{(#1)}}
\newcommand{\psub}[1]{{\cal P}_{{}_{\! \! #1}}}
\newcommand{\sst}[1]{{\scriptstyle #1}}
\newcommand{\ssst}[1]{{\scriptscriptstyle #1}}
\newcommand{\aft}{{{\scriptscriptstyle\succ}}}
\newcommand{\bef}{{{\scriptscriptstyle\prec}}}

\newcommand{\trans}[1]{\!{\xrightarrow{#1}}}

%%%%%%%%%%%%%%%%%%%%%%%%%%%%%%%%%%%%%%%%%%%%%%%%%%%%%%%%%%%%%%%%%%%%%%%%
%%%%%%%%%%%%%%%%%%%%%%%%%%%%%%%%%%%%%%%%%%%%%%%%%%%%%%%%%%%%%%%%%%%%%%%%

\title[Mean king]
{The ``mean king's problem" with continuous variables}

\author{  Alonso Botero }
\email{abotero@uniandes.edu.co} \affiliation{
    Departamento de F\'{\i}sica,
    Universidad de Los Andes,
    Bogot\'a, Colombia }\affiliation{ Department of Physics and Astronomy, University of
South Carolina, Columbia, SC 29208 }
\author{  Yakir Aharonov }
\affiliation{ Department of Physics and Astronomy, George Mason
University, Fairfax, VA 22030 }

\pacs{PACS numbers o3.65.Ta, 03.65.Ud, 03.67.-a}
\date{\today}

\begin{abstract}
\bigskip
We present the solution to  the ``mean king's problem" in the
continuous variable setting. We show that in this setting, the
outcome of a randomly-selected projective measurement of any
linear combination of the canonical variables $(\hat{x},\hat{p})$
can be ascertained with arbitrary precision. Moreover, we show
that the solution is in turn a solution to an associated
``conjunctive" version of the problem, unique to continuous
variables, where the inference task is to ascertain all the joint
outcomes of a simultaneous measurement of any number of linear
combinations of $(\hat{x},\hat{p})$.
\end{abstract}

\maketitle

A hallmark of the emergence of quantum information  has been the
shift in foundational research  towards probing interpretational
aspects of quantum mechanics  through information-theoretic tasks,
the feasibility of which is to be decided  within standard quantum
theory. The existence or non-existence of solutions to these tasks
has proved invaluable not only in revealing unsuspected
implications of quantum mechanics, but also in refining our
general heuristics of the theory, a good part of which inherits
from the early complementarity arguments of the ``old" quantum
theory.

One such task  is what has come to be known as the ``mean king's
problem". In its original version~\cite{VAA87} (and as retold in
the whimsical setting of Refs. \cite{AEng1sp1,Klap05}), a
physicist, Alice, is challenged by a mean king to precisely
ascertain the outcome of an ideal measurement that the king
performs of a  spin-1/2 observable randomly chosen from the
mutually complementary set $\{\hat{\sigma}_x, \hat{\sigma}_y,
\hat{\sigma}_z\}$. The conditions are that Alice can access the
system both before and after the king's measurement, and that only
at the end will the king reveal  the observable that was actually
measured and summon her to ascertain  the corresponding
measurement outcome.  While such a task stands in overt defiance
of the   complementarity heuristics, it was nevertheless
 shown in~\cite{VAA87} that the task is indeed feasible
according to quantum theory, with the solution involving initial
and final measurements by Alice performed on a composite system of
the spin and some other quantum particle. The extension of the
problem to arbitrary-dimensional discrete Hilbert spaces has also
been solved; first, for prime dimensions~\cite{EngA01}, and later,
through a connection to mutually-unbiased bases (MUB)
\cite{WoottFields}, for prime-power dimensions~\cite{Arav03}. More
recently, the problem was solved for  arbitrary discrete
dimensions~\cite{KimTanOz06} with  generalized (POVM) measurements
by Alice.

In this letter we present the solution to the mean king's problem
for the continuous variable (infinite-dimensional Hilbert space)
case, in which  Alice is challenged to infer the outcome from a
randomly chosen element of a continuous set of mutually
complementary measurements. The solution we present here is
derived from the solution to a closely-related joint measurement
inference problem addressed  by one of us in a recent publication
\cite{Botero07}. Consequently, the mean king's problem  can be
thought of as  dual to a related inference problem that is unique
to the continuous variable setting, and which involves the
simultaneous measurements of several complementary operators.  We
term this dual inference task  the ``conjunctive" version of the
mean king's problem, and show that the solution of the standard
king's problem is at the same time the solution of its associated
conjunctive version.

We begin by stating the continuous variable version of the
standard king's problem. Here, the king challenges Alice to
ascertain the outcome of an ideal measurement that he performs of
a quadrature observable
\begin{equation}
\hat{X}_\phi = \hat{x} \cos(\phi) + \hat{p}\sin(\phi)\, ,
\end{equation}
where $\hat{x}$ and $\hat{p}$ are canonically-conjugate operators
and $\phi$ is a randomly chosen angle from the interval $[0,
\pi)$. More generally, the king can perform any projective
measurement that distinguishes the elements of the   eigenbasis of
$\hat{X}_\phi$. Let $\ket{\xi}_\phi$  then be an eigenvector of
$\hat{X}_\phi$, with eigenvalue $\xi$, and distinguish the special
case of $\hat{X}_0 = \hat{x}$  with the notation $\ket{\xi}_x =
\ket{\xi}_0$. The eigenvectors of $\hat{X}_\phi$ are then given by
\begin{equation}\label{rdef}
\ket{\xi}_\phi = \hat{R}(\phi)\ket{\xi}_x\, , \ \ {\rm with } \ \
\hat{R}(\phi)\equiv e^{-i \frac{\phi}{2}(\hat{p}^2 + \hat{x}^2) }
\, .
\end{equation}
Alice's challenge is then to perform initial and final
measurements yielding outcomes  such that for every element of the
continuous family
 of bases $\{ \mathcal{B}(\phi)| \phi\in[0,\pi)\}$, where
$\mathcal{B}(\phi) = \{\ket{\xi}_\phi| \xi \in \mathbb{R}\}$, a
certain value $X_\phi$ is guaranteed to have been the result of
the king's measurement. In other words, Alice should be able to
assign a probability distribution $P(\xi) = \delta(\xi - X_\phi)$
for some $X_\phi$  to each $\mathcal{B}(\phi)$ measurement, based
on her measurement outcomes. Note that as in the discrete variable
case, the set of possible king's measurements involves mutually
complementary observables, or in other words, the basis set
$\{\mathcal{B}(\phi)\}$ consists of mutually unbiased bases: for
any two vectors from two bases $\mathcal{B}(\phi)$ and
$\mathcal{B}(\phi')$, we have
\begin{equation}
{}_{\phi'}\amp{\xi'}{\xi}_{\phi} =
{}_x\bra{\xi'}\hat{R}(\phi-\phi')\ket{\xi}_x\, .
\end{equation}
The matrix element is the Feynman propagator for a harmonic
oscillator of unit angular frequency, for a time interval
$\phi-\phi'$.  Since the modulus of the propagator of any
quadratic Hamiltonian is independent of the initial and final
coordinates\cite{FeyHibbs}, the modulus
$|{}_{\phi'}\amp{\xi'}{\xi}_{\phi}|$ is independent of $\xi$ and
$\xi'$.

A solution to the mean king's problem for the basis set $\{
\mathcal{B}(\phi) \}$ can then be obtained as follows: Alice will
have at her disposal another particle  with canonical
 variables $\hat{x}'$ and $\hat{p}'$. For the
two-particle system, we can define  the conjugate pairs
 $(\hatxpm,\hatppm)$
\begin{subequations}
\begin{eqnarray}\label{transeq}
\hatxplus = (\hat{x} + \hat{x}')/2 \, , & \ \ \ & \hatpplus =
\hat{p}+
\hat{p}' \, ,  \\
\hatxmin = \hat{x} - \hat{x}' \,  , & \ \ \ & \hatpmin = (\hat{p}
-\hat{p}')/2 \, .
\end{eqnarray}
\end{subequations}
In particular, she  will need to perform initial and final
measurements  yielding  eigenstates of the commuting pair
$(\hat{x}_+,\hat{p}_-)$, which up to normalization are given by
\begin{equation}
\ket{\xplus,\pmin} = \int_{-\infty}^{\infty}\! ds\, e^{ i s \pmin}
\left|\left.\xplus+\frac{s}{2}\right\rangle_{\!\!x}\right.
\left|\left.\xplus-\frac{s}{2}\right\rangle_{\!\!x'}\right. \, ,
\end{equation}
and  eigenstates of the commuting pair $(\hat{x}_-,\hat{p}_+)$,
given by
\begin{equation}
\ket{\xmin,\pplus} = \int_{-\infty}^{\infty}\! ds\, e^{i s \pplus}
\left|\left.s+\frac{\xmin}{2}\right\rangle_{\!\!x}\right.
\left|\left.s-\frac{\xmin}{2}\right\rangle_{\!\!x'}\right. \, .
\end{equation}
For later convenience, we define the phase factor $e^{i \gamma}
\equiv \amp{\xmin,\pplus}{\xplus,\pmin} = e^{i (\xplus \pplus -
\xmin \pmin)}$. Given the projector $\hat{\Pi}_\phi(\xi) =
\ket{\xi}_\phi\bra{\xi}$, the probability amplitude
$\mel{\xmin,\pplus}{\hat{\Pi}_\phi(\xi)}{\xplus,\pmin}$ can then
be written as
\begin{equation}
e^{-i \gamma}\mel{\xmin,\pplus}{\hat{\Pi}_\phi(\xi)}{\xplus,\pmin}
=  {}_{\phi}\mel{\xi}{\hat{\Delta}(x,p)}{\xi}_\phi ,
\end{equation}
where  we define the labels
\begin{equation}
\label{xpvals} x \equiv \xplus + \xmin/2 \, , \ \ \ \  p \equiv
\pmin + \pplus/2\, ,
\end{equation}
and the operator
\begin{eqnarray}
\hat{\Delta}(x,p) & = & e^{- i \gamma} {\rm Tr}'\left(
\ket{\xplus,\pmin}\bra{\xmin,\pplus} \right)\nonumber \\ & = &
\int_{-\infty}^{\infty}\! ds\, e^{ i s p}
\left|\left.x+\frac{s}{2}\right\rangle_{\!\!x}\right.
\left.\left\langle x-\frac{s}{2}\right.\right|_{x}\, ,
\end{eqnarray}
with ${\rm Tr'}(\ldots)$ denoting traces with respect to the
primed degree of freedom. The operator $\hat{\Delta}(x,p)$ is
the
 Weyl-symmetrized bivariate $\delta$-function\cite{Leaf68},
\begin{equation}
 \hat{\Delta}(x,p) =\frac{1}{2
\pi}\int_{\mathbb{R}^2}\! d^2 \vect{\chi}\, e^{ -i\left(
\chi_1(x-\hat{x}) + \chi_2(p-\hat{p}) \right)}\, ,
\end{equation}
and has a Wigner function that is a $\delta$-function in
phase-space. However, being non-positive and non-local in the
$\hat{x}$ representation, $\hat{\Delta}(x,p)$ is  more accurately
interpreted as a generalized parity operator\cite{Oz98} (up to a
factor).   The connection with the Wigner representation
nevertheless proves useful for calculations.  In particular, we
find  for the case $\phi=0$, the expectation value
\begin{equation}
\qave{\xi}{\hat{\Delta}(x,p) }_x = \delta(\xi-x) \, .
\end{equation}
Similarly, for the general case $\phi\neq 0$, we note that
$\hat{R}(\phi)$ in Eq.~(\ref{rdef}) implements a rotation of the
canonical operators, and in turn  a rotation of the labels of
$\hat{\Delta}(x,p)$ through
\begin{equation} \hat{R}^\dagger(\phi)
\hat{\Delta}(x,p)\hat{R}(\phi) = \hat{\Delta}(X_\phi,Y_\phi)\, ,
\end{equation}
where the rotated labels $(X_\phi,Y_\phi)$ are the c-number
quadratures obtained from $(x,p)$
\begin{subequations}
\label{quadvals}
\begin{eqnarray}
X_\phi & = & x \cos\phi + p \sin
\phi\, , \label{quadvalsx}\\
Y_\phi & =  &p\cos \phi -x \sin\phi\, . \label{quadvalsy}
\end{eqnarray}
\end{subequations}
It  therefore follows that for all $\phi$,
\begin{equation}\label{impropamp}
\mel{\xmin,\pplus}{\hat{\Pi}_{\phi}(\xi)}{\xplus,\pmin} = e^{i
\gamma}  \delta(\xi-X_\phi) \, ,
\end{equation}
and hence the amplitude vanishes unless $\xi = X_\phi$. We note
that this result is the natural extension to the continuum of a
known connection between the  king's problem and discrete Weyl and
Wigner distributions~\cite{Durt04}.

Finally, for a rigorous statement of probabilities in the
continuous variable case, we need to describe the king's
projective measurement as the limit of a sequence of POVM
measurements. Let the POVM $\{\hat{E}(\xi|\phi,\epsilon)|\xi \in
\mathbb{R}\}$ be any operator-valued $\delta$-sequence,
\begin{equation}
\int_{-\infty}^{\infty}\!d\xi\, \hat{E}(\xi|\phi,\epsilon) =
\openone \, , \ \ \ \lim_{\epsilon\rightarrow
0}\hat{E}(\xi|\phi,\epsilon) = \delta(\xi - \hat{X}_\phi) \, ,
\end{equation}
admitting a $\mathcal{B}(\phi)$-diagonal Kraus decomposition $
\hat{E}(\xi|\phi,\epsilon) = \sum_i
\hat{A}^\dagger_i(\xi|\phi,\epsilon)\hat{A}_i(\xi|\phi,\epsilon) $
with
\begin{equation}
\hat{A}_i(\xi|\phi,\epsilon) = \int_{-\infty}^{\infty}\!d\xi'
a_i(\xi,\xi'|\phi,\epsilon)\hat{\Pi}_{\phi}(\xi')\, .
\end{equation}
When conditioned on initial and final states, the outcome
distribution of the POVM measurement is then given by
\begin{eqnarray}
P(\xi|\phi,\epsilon) =
\frac{\sum_i|\mel{\xmin,\pplus}{\hat{A}_i(\xi|\phi,\epsilon)}{\xplus,\pmin}|^2}
{\sum_i\int_{-\infty}^{\infty}\!d\xi\,
|\mel{\xmin,\pplus}{\hat{A}_i(\xi|\phi,\epsilon)}{\xplus,\pmin}|^2}
\, .
\end{eqnarray}
Using  Eq.~(\ref{impropamp}),  we find the amplitudes
\begin{equation}
|\mel{\xmin,\pplus}{\hat{A}_i(\xi|\phi,\epsilon)}{\xplus,\pmin}|^2
= |a_i(\xi,X_\phi|\phi,\epsilon)|^2 \, ,
\end{equation}
and hence the outcome   distribution
\begin{equation}
P(\xi|\phi,\epsilon) =
Z^{-1}\sum_i|a_i(\xi,X_\phi|\phi,\epsilon)|^2\, ,
\end{equation}  where $Z$ is a normalization constant. Then, from the
condition $\lim_{\epsilon\rightarrow 0}\hat{E}(\xi|\phi,\epsilon)
= \delta(\xi - \hat{X}_\phi)$ and the diagonality of the Kraus
operators, the condition $\lim_{\epsilon\rightarrow 0} \sum_i
|a_i(\xi,X_\phi|\phi,\epsilon)|^2 = \delta(\xi-X_\phi) $ follows.
Thus,
\begin{equation}
\lim_{\epsilon\rightarrow 0}P(\xi|\phi,\epsilon)=
 \delta(\xi-X_\phi)\, ,
\end{equation}
which is the desired result. So indeed, if Alice makes  initial
and final projective measurements of the commuting pairs
$(\hatxplus,\hatpmin)$ and $(\hatxmin,\hatpplus)$ respectively,
she will be able to retrodict the outcome $X_\phi$ of the king's
intermediate measurement of any $\hat{X}_\phi$ for all values of
$\phi$, in the limit of an infinitely sharp measurement.

We have thus seen how, in conformity with the
 standard formulation of the king's problem,
a value $X_\phi$ is assigned to every possible $\hat{X}_\phi$
 in a context of
{\em exclusive disjunction}; in other words, the assignment makes
reference to a single intermediate measurement that is performed
by the king at the exclusion of the other possibilities. One may
then enquire as to whether there exists a ``dual" measurement
context consistent with the assignment of a value $X_\phi$ to
every  $\hat{X}_\phi$, but in {\em conjunction}.
 This situation was
partially probed in Ref.~\cite{Botero07}, where the case of two
incompatible dense observables was considered. The results,
however, have a natural extension to any number of linear
combinations of the two observables.  This extension therefore
defines the dual measurement context for the continuous variable
king's problem--what we henceforth term the  ``conjunctive"
version of the  problem.

The conjunctive version of the mean king's problem has to do with
what at first sight would appear to be a truly impossible (if not
ill-defined) inference task according to the complementarity
heuristics.  In this case, the king measures, simultaneously and
\emph{sharply}, a whole set of quadrature operators
$\hat{X}_{\phi_1}, \hat{X}_{\phi_2}, \ldots \hat{X}_{\phi_n}$,
with an operational definition of such a sharp measurement
procedure to be given shortly. Alice is again put to the task of
guessing the outcomes of the measurement, and with the same
initial and final measurements discussed earlier, we shall see
that she is able to retrodict all the outcomes of the king's
single joint measurement.

We define  the simultaneous sharp  measurement (SSM) of a set of
non-commuting quadrature operators $\hat{\vect{X}} =
(\hat{X}_{\phi_1},\hat{X}_{\phi_2}, \ldots,\hat{X}_{\phi_n})$ in
terms of a POVM measurement with continuous index set
$\vect{\xi}=( {\xi}_1, {\xi}_2,\ldots {\xi}_n)$ representing the
measurement outcomes.  The POVM is assumed to admit a Kraus
decomposition  $ \hat{E}(\vect{\xi}|\epsilon) = \sum_i
\hat{A}_i^\dagger(\vect{\xi}|\epsilon)\hat{A}_i(\vect{\xi}|\epsilon)
$ with Kraus operators of the form
\begin{equation}
\hat{A}_i(\vect{\xi}|\epsilon) =\,
:f_{i}(\vect{\xi}-\hat{\vect{X}}|\epsilon): \, ,
\end{equation}
where  $:\, \ldots\, :$ stands for the generalized Weyl-symmetric
ordering
\begin{equation}\label{Weylsym}
:f_i(\vect{\xi}-\hat{\vect{X}}|\epsilon): = \frac{1}{(2
\pi)^n}\!\int_{\mathbb{R}^n}\!\!\! \!d^n\!\vect{\xi}'\!\!
\int_{\mathbb{R}^n}\!\!\!\!d^n\!\vect{\chi}\, e^{ -i
\vect{\chi}\cdot(\vect{\xi} - \hat{\vect{X}}-\vect{\xi}'
)}f_i({\vect{\xi}'}|\epsilon) \, ,
\end{equation}
and where the $f_i({\vect{\xi}}|\epsilon)$ are functions defining
an $n$-dimensional $\delta$-sequence in the sense that
\begin{equation}\label{ndeltseq}
 \lim_{\epsilon\rightarrow
0}\sum_i|f_i({\vect{\xi}}|\epsilon)|^2 =
\delta^{(n)}({\vect{\xi}})\, .
\end{equation}
This choice of POVM arises naturally from a measurement
implemented by $n$ canonical  instruments described by some
initial state $\hat{\rho}$, with  pointer variables
$\hat{\vect{\xi}}$, and simultaneously coupled to the system
through an impulsive Hamiltonian
\begin{equation}
\hat{H} = -\delta(t) \hat{\vect{\chi}}\cdot\hat{\vect{X}}\, ,
\end{equation}
where the $\hat{\chi}_i$ are conjugate to the pointer variables
$([\hat{\chi}_i,\hat{\xi}_j]=i \delta_{ij}$). With an arbitrary
decomposition of $\hat{\rho} = \sum_i p_i
\ket{\phi_i}\bra{\phi_i}$, the Kraus operators are then given by
\begin{equation}
\hat{A}_i(\vect{\xi}|\epsilon) = \sqrt{p_i}\bra{\vect{\xi}}e^{i
\hat{\vect{\chi}}\cdot\hat{\vect{X}}}\ket{\phi_i}\, .
\end{equation}
``Sharpness" in this context therefore means that the
$\delta$-sequence is defined from $\langle
\vect{\xi}|\hat{\rho}|\vect{\xi}\rangle =
\sum_i|f_i({\vect{\xi}}|\epsilon)|^2$, the pre-measurement
probability distribution of the pointer variables, as in the case
of a projective measurement.

Now, in accordance with the uncertainty principle, the mutual
back-reaction between the different instruments ensures that the
probability distribution  of the final readings from any
preselected ensemble will necessarily be unsharp~\cite{ArtGoo88}.
To see this, note that the SSM can be described in the Heisenberg
picture as the action of the operator $ \hat{U} = e^{i
\hat{\vect{\chi}}\cdot\hat{\vect{X}} } $ on the instrument pointer
variables $\hat{\vect{\xi}}$.  The pointer operators representing
the final outcome can then be written as
\begin{equation}
\hat{\vect{\xi}} = \hat{U}^\dagger\hat{\vect{\xi}}\hat{U}|_{\rm in} \\
= \hat{\vect{\xi}}_{\rm in} + \hat{\vect{X}}_{\rm in} +
\frac{1}{2}\vect{C}\hat{\vect{\chi}}_{\rm in}
\end{equation}
 where ``${\rm in}$" denotes the pre-measurement Heisenberg variables and  where
 $\vect{C}$ is the antisymmetric matrix with elements
\begin{equation}
C_{ij} = i[
\hat{X}_{\phi_i},\hat{X}_{\phi_j}]=\sin(\phi_i-\phi_j)\, ,
\end{equation}
arising from the commutator of the quadrature operators. The
back-reaction of the instruments is reflected in a dependence of
the outcomes $\hat{\vect{\xi}}$ on the conjugate variables
$\hat{\vect{\chi}}_{\rm in}$, which are uncertain to the same
extent that the initial pointer variables $\hat{\vect{\xi}}_{\rm
in}$ are certain. In particular, in the ``sharp" limit
$\epsilon\rightarrow 0$, the uncertainties in the readings from
any preselected ensemble are  dominated by the divergent
uncertainties of the conjugate $\hat{\vect{\chi}}_{\rm in}$
variables. The prospect of any precise inference of the joint
outcomes would therefore seem to be doomed based on the mutual
disturbance of the simultaneous measurements.

Let us, however, compute the posterior probability distribution of
outcomes of the SSM, when conditioned on the initial and final
states previously discussed. The probability distribution is then
given by
\begin{equation}
P(\vect{\xi}|\epsilon) = Z^{-1}\sum_i \left|
\mel{\xmin,\pplus}{\hat{A}_i(\vect{\xi}|\epsilon)}{\xplus,\pmin}\right|^2.
\end{equation}
Given the definition of ${\hat{A}_i(\vect{\xi}|\epsilon)}$ in
terms of the Weyl-symmetrization prescription,~(\ref{Weylsym}),
the amplitude
$\mel{\xmin,\pplus}{\hat{A}_i(\vect{\xi}|\epsilon)}{\xplus,\pmin}$
is the convolution of $f_i(\vect{\xi}|\epsilon)$ and the Fourier
transform of
\begin{equation}
\mel{\xmin,\pplus}{e^{ i \vect{\chi}\cdot
\hat{\vect{X}}}}{\xplus,\pmin}\, .
\end{equation}
Concentrating on this matrix element, we write  the exponent
$\vect{\chi}\cdot \hat{\vect{X}}$ as
\begin{equation}
\vect{\chi}\cdot \hat{\vect{X}} = a \hat{x} + b \hat{p}
\end{equation}
for some $\vect{\chi}$-dependent $a$ and $b$. Next we note that
the linear combination $a \hat{x} + b \hat{p}$ can also be written
 in terms of the collective canonical variables $(\hat{x}_\pm,
\hat{p}_\pm)$ as
\begin{equation}
a \hat{x} + b \hat{p} = \hat{L}(a,b) + \hat{R}(a,b)\, ,
\end{equation}
where $\hat{L}(a,b)$ and $ \hat{R}(a,b)$ are \emph{commuting}
operators defined as
\begin{eqnarray}
\hat{L}(a,b) & = & (a \hatxmin + b \hatpplus)/2\, , \\
 \hat{R}(a,b) & = & a \hatxplus + b \hatpmin \, .
\end{eqnarray}
Consequently, the operator exponential can be factored as
\begin{equation}
e^{i(a \hat{x} + b \hat{p})} = e^{i \hat{L}(a,b)}e^{i
\hat{R}(a,b)}\, .
\end{equation}
Since the eigenvectors $\ket{\xmin,\pplus}$ and
$\ket{\xplus,\pmin}$ are respectively eigenvectors of
$\hat{L}(a,b)$ and $\hat{R}(a,b)$, we then verify that
\begin{eqnarray}
\mel{\xmin,\pplus}{e^{i(a \hat{x} + b \hat{p})} }{\xplus,\pmin} =
e^{i \gamma } e^{i (a x + i b p)}\, ,
\end{eqnarray}
where $x$ and $p$ are as defined in Eq.~(\ref{xpvals}), or
equivalently,
\begin{equation}
\mel{\xmin,\pplus}{e^{ i \vect{\chi}\cdot
\hat{\vect{X}}}}{\xplus,\pmin} = e^{i \gamma }e^{ i
\vect{\chi}\cdot {\vect{X}}}\, ,
\end{equation}
where ${\vect{X}}= (X_{\phi_1},X_{\phi_1},\ldots,X_{\phi_n})$ with
the $X_{\phi}$  as defined in Eq.~(\ref{quadvalsx}). The Fourier
transform of the matrix element is therefore a $\delta$-function
at $\vect{X}$, so that upon convolution,   we find that
\begin{equation}
\mel{\xmin,\pplus}{\hat{A}_i(\vect{\xi}|\epsilon)}{\xplus,\pmin} =
e^{i \gamma } f_i(\vect{\xi}- \vect{X}|\epsilon)\, .
\end{equation}
 The joint outcome
probability distribution for the measurement  is then
\begin{equation} P(\vect{\xi}|\epsilon) =
\sum_i|f_i(\vect{\xi}-\vect{X}|\epsilon)|^2\, , \end{equation} and
with the $\delta$-sequence condition~(\ref{ndeltseq}), we finally
obtain
\begin{equation}
\lim_{\epsilon\rightarrow 0}P(\vect{\xi}|\epsilon) =
\delta^{(n)}(\vect{\xi}-\vect{X})\, .
\end{equation}
Alice is therefore able to ascertain the king's joint measurement
outcomes in the sharp limit of the conjunctive version of the
challenge, as originally advertised.

The exact coincidence between the conditional outcomes of the
conjunctive version of the king's problem and the outcomes in the
standard version would seem to suggest that Alice's assignments
are, in some sense, non-contextual (independent of measurement
conditions). However,  as in  other situations where conditioning
from postselection allows precise assignments to incompatible
measurement outcomes~\cite{Mermin95,LeifSpekk05}, it is also
possible to find for the same pre and postselections,  potential
intermediate measurements exhibiting contextuality in the present
case. An interesting example highlighting the role of entanglement
has to do with the intermediate SSM measurement  of two sets of
quadrature-operator arrays, $\hat{\vect{X}}$ and
$\hat{\vect{X}}'$, the second set involving operators of Alice's
ancillary particle. By the symmetry of the problem, sharp
conditional probabilities can also be assigned to the SSM
measurement outcomes of $\hat{\vect{X}}'$, but only in the absence
of measurements on the original particle. On the other hand (and
as discussed briefly in~\cite{Botero07}), a non-sharp conditional
probability distribution  generally follows for the full outcome
set in the SSM of $\hat{\vect{X}}$ {\em and} $\hat{\vect{X}}'$,
showing that the assignment of sharp values to either
$\hat{\vect{X}}$ or $\hat{\vect{X}}'$ is, in fact, contextual.
Nonetheless, our results show that entanglement allows for a
surprising degree of flexibility in the composition of the
observable sets that reveal the contextual aspects of retrodiction
from postselection.

\end{document}